\documentclass[10pt]{article}
\usepackage{amssymb,amsmath,amsthm}
\usepackage{mathrsfs}
\newcommand{\ie}{\emph{i.e.}}
\newcommand{\eg}{\emph{e.g.}}
\newcommand{\cf}{\emph{cf}}
\newcommand{\ad}{\emph{ad}}
\newcommand{\etc}{\emph{etc}}
\newcommand{\demi}{\frac{1}{2}}
\newcommand{\Sphere}{\mathbb{S}}

\newcommand{\Real}{\mathbb{R}}
\newcommand{\Nat}{\mathbb{N}}

\newcommand{\Smooth}{C}

\newcommand{\sii}{L^2}
\newcommand{\sinf}{L^\infty}
\newcommand{\sobi}{\mathop{W_0^{1,2}}\nolimits}
\newcommand{\Sobi}{\mathop{W^{1,2}}\nolimits}

\newcommand{\Dom}{\mathop{\mathrm{Dom}}\nolimits}

\newcommand{\diag}{\mathop{\mathrm{diag}}\nolimits}
\newcommand{\Id}{1}
\newcommand{\tube}{\Omega}
\newcommand{\curve}{\Gamma}
\newcommand{\curvature}{\mathcal{K}}

\newcommand{\cross}{\omega}
\newcommand{\Circle}{\mathcal{C}}
\newcommand{\vol}{\mathrm{vol}}
\newcommand{\tubemap}{\mathcal{L}}
\newcommand{\tubemapalt}{\mathscr{L}}
\newcommand{\Hilbert}{\mathcal{H}}
\newcommand{\both}{\iota}

\newcommand{\HFrenet}{\langle\mathrm{H1}\rangle}
\newcommand{\HBasic}{\langle\mathrm{H2}\rangle}
\newenvironment{Assumption}[1]
{\begin{description}\item[$#1$]\it}
{\end{description}}
\newtheorem{Theorem}{Theorem}
\newtheorem{Lemma}{Lemma}
\newtheorem{Proposition}{Proposition}

\theoremstyle{definition}

\newtheorem{remark}{remark}
\newtheorem{Remark}[remark]{Remark}
\theoremstyle{remark}

\def\OMIT#1{}
\begin{document}
%
%
\title{{\Large\textbf{
A lower bound to the spectral threshold in curved tubes
}}
}
\author{
P.~Exner$^{1,2}$,
P.~Freitas$^3$
and D.~Krej\v{c}i\v{r}\'{\i}k$^{1,3}$
}
\date{\footnotesize
\begin{quote}
\emph{
\begin{itemize}
\item[$^1$]
Department of Theoretical Physics, Nuclear Physics Institute, \\
Academy of Sciences, 250\,68 \v{R}e\v{z} near Prague, Czech Republic
\item[$^2$]
Doppler Institute, Czech Technical University, \\
B\v{r}ehov{\'a}~7, 115\,19 Prague, Czech Republic
\item[$^3$]
Departamento de Matem\'atica, Instituto Superior T\'ecnico, \\
Av. Rovisco Pais, 1049-001 Lisboa, Portugal
\item[]
\emph{E-mail:}
exner@ujf.cas.cz,
pfreitas@math.ist.utl.pt,
dkrej@math.ist.utl.pt
\end{itemize}
}
\end{quote}
28 April 2004
}
\maketitle
%
%
\begin{abstract}
\noindent  
We consider the Laplacian in curved tubes of arbitrary
cross-section rotating together with the Frenet frame 
along curves in Euclidean spaces of arbitrary
dimension, subject to Dirichlet boundary conditions on the
cylindrical surface and Neumann conditions at the ends of the tube.
We prove that the spectral threshold of the Laplacian is estimated
from below by the lowest eigenvalue of the Dirichlet Laplacian in
a torus determined by the geometry of the tube.
\end{abstract}
%
%
%

\section{Introduction}
%
Problems linking the shape of a region to the spectrum of the
associated Laplacian, subject to various boundary conditions, have
been considered for more than a century. While classical
motivations came from theories of elasticity, acoustics,
electromagnetism, \etc, in the quantum-mechanical context a strong
fresh impetus is mostly due to the recent technological progress
in semiconductor physics.

More specifically, the Dirichlet Laplacian in infinite plane
strips or space tubes of constant cross-section is widely used as
a mathematical model for the Hamiltonian of a quantum particle in
mesoscopic structures called \emph{quantum waveguides}
\cite{DE,LCM,Hurt}. The existence of geometrically induced
\emph{bound states} in curved asymptotically straight waveguides
is probably the most interesting theoretical result for these
systems \cite{ES,GJ,RB,DE,KKriz,ChDFK}. Indeed, these bound
states, which are known to perturb the particle transport, are of
pure quantum origin because there are no classical closed
trajectories in the tubes in question, apart from a zero measure
set of initial conditions in the phase space. Mathematically, one
deals with the discrete spectrum of the Dirichlet Laplacian, which
is a non-trivial property for unbounded regions. The principal
objective of this paper is to establish a lower bound to the
ground-state energies of curved quantum waveguides.

We proceed in greater generality by considering $d$-dimensional
tubes, unbounded or bounded, with any $d \geq 2$ and arbitrary
cross-section rotating along a reference curve together with the Frenet frame. 
At the same time, we do not restrict ourselves to
asymptotically straight tubes, \ie, if the tube is unbounded, the
estimated spectral threshold of the Laplacian may not be a
discrete eigenvalue, but rather the threshold of the essential spectrum;
this happens, for instance, if the tube is periodically curved.

To state the main result of the paper,
let us introduce some notation.
Given a bounded or unbounded open interval~$I$,
let~$\curve:I\to\Real^d$ be a unit-speed curve
with curvatures~$\kappa_i:I\to\Real$, $i\in\{1,\dots,d-1\}$,
w.r.t.\ an appropriate smooth Frenet frame $\{e_1,\dots,e_d\}$,
\cf~the assumption~$\HFrenet$ below.
Given a bounded open connected set $\cross\in\Real^{d-1}$
with the centre of mass at the origin,
we define the tube~$\Omega$ by rotating~$\cross$ along the curve
together with the Frenet frame,
\ie,
\begin{equation}\label{tube}
  \Omega := \tubemap(I\times\cross) ,
  \qquad
  \tubemap(s,u_2,\dots,u_d):= \curve(s)+ e_\mu(s) \, u_\mu ,
\end{equation}
(the repeated indices convention is adopted throughout the paper,
the Latin and Greek indices run through $1,2,\dots,d$ and $2,\dots,d$,
respectively).
We make the assumption~$\HBasic$ below (\cf~Remark~\ref{Rem.basic})
in order to ensure that
$
  \tubemap:I\times\cross\to\tube
$
is a diffeomorphism.
Our object of interest is the non-negative Laplacian
\begin{equation}\label{Laplacian}
  -\Delta
  \qquad\textrm{on}\quad\ \sii(\Omega)
  \,,
\end{equation}
subject to Dirichlet boundary conditions
on the cylindrical part of the boundary~$\tubemap(I\times\partial\cross)$
and, if~$\partial I$ is not empty, Neumann boundary conditions 
on the remaining boundary~$\tubemap((\partial I)\times\cross)$.
Our main result reads as follows.
\begin{Theorem}\label{Thm.bound}
Suppose the assumptions $\HFrenet$ and $\HBasic$ are satisfied.
Then
\begin{equation}\label{bound}
  \inf\sigma(-\Delta) \geq
  \min\left\{
  \lambda_0(\sup\kappa_1),\lambda_0(\inf\kappa_1)
  \right\},
\end{equation}
where
$
  \lambda_0(\kappa) \geq c > 0
$ 
denotes the spectral threshold
of~$-\Delta$ in the tube of cross-section~$\cross$ built either
over a circle of curvature~$\kappa$ if $\kappa\not=0$ or over 
a straight line if~$\kappa=0$;
$c$ is a constant depending only on~$\cross$ and~$d$.
\end{Theorem}

The lower bound of Theorem~\ref{Thm.bound} holds, of course, 
for other boundary conditions imposed 
on~$\tubemap((\partial I)\times\cross)$, \cf~Section~\ref{Sec.Disc}.

Note that $\lambda_0(\kappa)$
is the lowest eigenvalue of the Dirichlet Laplacian
in a torus of cross-section~$\cross$ if~$\kappa\not=0$
or the threshold of the essential spectrum
of the Dirichlet Laplacian in an infinite straight tube
of cross-section~$\cross$
(which is the lowest eigenvalue~$\mu_0$
of the Dirichlet Laplacian in~$\cross$)
if~$\kappa=0$, \cf~Section~\ref{Sec.Torus}.
Thus the claim of Theorem~\ref{Thm.bound} can be expressed 
illustratively as follows: 
take an ``osculation torus" at each point of $\Gamma$
(\ie~the torus with the identical cross-section 
built over the osculation circle to $\Gamma$ at the point), 
then the bound~(\ref{bound}) corresponds 
to the smallest of this tori spectral thresholds. 
The uniform lower bound given by the geometric constant~$c$ 
is a consequence of the Faber-Krahn inequality, 
\cf~Proposition~\ref{Prop.FK}. 
 
We stress again that while the spectrum of~(\ref{Laplacian})
is purely discrete whenever~$I$ is bounded,
$\sigma(-\Delta)$ has in general both discrete
and essential parts in the unbounded case.
For instance, if~$I=\Real$, $\cross=B_a$ (ball of radius $a>0$),
$\kappa_1\not=0$ but $\kappa_1(s)\to 0$ as $|s|\to\infty$,
it is known from~\cite{ChDFK}
that $\sigma_\mathrm{ess}(-\Delta)=[\mu_0,\infty)$
and there are always discrete eigenvalues in $(0,\mu_0)$.

While bounds on the eigenvalues for the Laplacian on
bounded subsets of~$\Real^d$ have been studied by many authors
(see~\cite{Henrot_2003} for an overview), to the best of our
knowledge there is only one previous result on the lower bound to
the spectral threshold of the Laplacian in unbounded tubes. Using
the Payne-P\'olya-Weinberger conjecture \cite{PPW1,PPW2} proved
then in \cite{Ashbaugh-Benguria_1991}
(see also~\cite{Ashbaugh-Benguria_1992}), 
M.~S.~Ashbaugh and the first author 
derived in~\cite{AE} a lower bound in the situation
when~$I=\Real$, $d=2,3$, the cross-section was circular and the
discrete spectrum of~$-\Delta$ was not empty but finite. 
As we discuss at the end of Section~\ref{Sec.Disc}, 
our Theorem~\ref{Thm.bound} provides a better bound and applies to
tubes with an infinite number, or without any, discrete
eigenvalues, too. On the other hand, the approach of~\cite{AE}
applies to more general forms of~$\Omega$ than the regular tubes
considered here. Let us also mention that one can use the results
of~\cite{EW} to derive a Lieb-Thirring-type inequality
for~$-\Delta$.

The heuristic idea behind the proof of Theorem~\ref{Thm.bound} is as follows.
For a moment, let us assume that~$\kappa_1$ is piece-wise constant
and all~$\kappa_\mu=0$,
so that~$I$ is a closure of the union of~$L$ (possibly $L=\infty$)
open subintervals~$I_\ell$, $\ell\in\{1,\dots,L\}$,
and each~$\curve_\ell:=\curve(I_\ell)$
is a circular or straight segment.
We have
$
  -\Delta \geq \bigoplus_{\ell=1}^L (-\Delta^\ell) ,
$
where each $-\Delta^\ell$ is the Laplacian
on~$\sii(\tubemap(I_\ell\times\cross))$
with Dirichlet boundary conditions on~$\tubemap(I_\ell\times\partial\cross)$
and the Neumann ones on $\tubemap((\partial I_\ell)\times\cross)$.
Note that $\inf\sigma(-\Delta^\ell)$
does not depend on the length of~$\curve_\ell$
because the first (generalised) eigenfunction
of the Dirichlet Laplacian in a torus or an infinite straight tube
is invariant w.r.t.\ to rotations or translations, respectively.
Consequently, $\inf\sigma(-\Delta^\ell)=\lambda_0(\kappa_1^\ell)$,
where~$\kappa_1^\ell$ denotes the first curvature of~$\curve_\ell$.
The spectral threshold of $-\Delta$ is thus estimated from below
by $\min_\ell \lambda_0(\kappa_1^\ell)$
and an analysis of the properties of the first eigenvalue in the torus
(Section~\ref{Sec.Torus}) shows that this minimum is equal to
$
  \min\{\lambda_0(\max_\ell \kappa_1^\ell),\lambda_0(\min_\ell\kappa_1^\ell)\}
$
(note that $\kappa\mapsto\lambda_0(\kappa)$ may not be even
for a general cross-section~$\cross$).
An important consequence of 
(geometric) Lemma~\ref{Lemma.positive} below 
is that this lower bound is not affected 
by higher curvatures~$\kappa_\mu$. 
Then the general result of Theorem~\ref{Thm.bound}
follows by the above procedure
at once if one considers the Laplacian through its quadratic form
(because the supplementary Neumann conditions do not appear explicitly
in the form domain).

The organisation of the paper is as follows. The tube~$\Omega$ and
the corresponding Laplacian~$-\Delta$ are properly defined
in the preliminary Section~\ref{Sec.Preliminary}. 
In Section~\ref{Sec.Inter}, we prove the geometric Lemma~\ref{Lemma.positive}
and an intermediate lower bound, Theorem~\ref{Thm.bound.begin}, 
as its direct consequence. 
Theorem~\ref{Thm.bound} then immediately follows 
from Theorem~\ref{Thm.bound.begin} 
and results in Section~\ref{Sec.Torus},
which is devoted to a detailed analysis of spectral properties of~$-\Delta$ 
in the case where the reference curve~$\curve$ is a circular segment.   
Finally, in Section~\ref{Sec.Disc},
we summarise the results obtained,
discuss possible extensions and refer to some open problems.
We conclude the paper by comparing our result with the lower bound
found in~\cite{AE} for a special case of infinite tubes
in two and three dimensions. 

\section{Preliminaries}\label{Sec.Preliminary}
%
\subsection{The reference curve}
Given an open interval $I \subseteq \Real$
and an integer $d \geq 2$,
let $\curve: I \to \Real^d$ be a unit-speed $\Smooth^{d-1}$-smooth curve
satisfying
\begin{Assumption}{\HFrenet}
$\curve$ possesses a positively oriented
$\Smooth^1$-smooth Frenet frame $\{e_1,\dots,e_d\}$
with the properties that
$
  e_1=\dot{\curve}
$
and
$$
  \forall i \in \{1,\dots,d-1\}, \
  \forall s\in I,
  \quad
  \dot{e}_i(s) \
  \ \mbox{lies in the span of} \
  e_1(s),\dots,e_{i+1}(s).
$$
\end{Assumption}
\begin{Remark}\label{Rem.Frenet}
We refer to~\cite[Sec.~1.2]{Kli} for the notion of Frenet frames.
A sufficient condition to ensure the existence
of the Frenet frame of $\HFrenet$
is to require that for all $s \in \Real$, the vectors
$
  \dot{\curve}(s), \curve^{(2)}(s), \dots, \curve^{(d-1)}(s)
$
are linearly independent, \cf~\cite[Prop.~1.2.2]{Kli}.
This is always satisfied if $d=2$.
However, we prefer not to assume \emph{a priori}
this non-degeneracy condition for $d \geq 3$ because then
one excludes the curves such that $\curve\upharpoonright I_1$
lies in a lower-dimensional subspace of~$\Real^d$
for some open $I_1 \subseteq I$.
Further comments on the assumption~$\HFrenet$ 
will be given in the closing section.
\end{Remark}
\noindent
We have the Serret-Frenet formulae,
\cf~\cite[Sec.~1.3]{Kli},
\begin{equation}\label{Frenet}
  \dot{e}_i = \curvature_{ij} \, e_j
\end{equation}
where $\curvature\equiv(\curvature_{ij})$
is the skew-symmetric $d \times d$ matrix defined by
\begin{equation}\label{curvature}
  \curvature :=
  \begin{pmatrix}
   0                & \kappa_1 &               & \textrm{\LARGE 0}\\
   -\kappa_1        & \ddots   & \ddots        &                 \\
                    & \ddots   & \ddots        & \kappa_{d-1}    \\
   \textrm{\LARGE 0} &          & -\kappa_{d-1} & 0
   \end{pmatrix}.
\end{equation}
Here~$\kappa_i$ is called the $i^\mathrm{th}$ curvature of~$\curve$
which is, under our assumptions,
a continuous function of the arc-length parameter~$s \in I$.

\subsection{Tubes}
Let~$\cross$ be an arbitrary bounded open connected set in~$\Real^{d-1}$.
Without loss of generality, we assume that~$\cross$
is translated so that its centre of mass is at the origin.
Put $\tube_0:=I \times \cross$ and $u:=(u_2,\dots,u_d)\in\cross$.
We define the tube~$\tube$ built over~$\curve$
as the image of the mapping $\tubemap: \tube_0 \to \Real^d$
defined in~(\ref{tube}), \ie~$\tube:=\tubemap(\tube_0)$.
Assuming that
\begin{equation}\label{Ass.Basic}
  \tubemapalt:\tube_0\to\tube:\{(s,u)\mapsto\tubemap(s,u)\}
  \quad
  \mbox{is a $\Smooth^1$-diffeomorphism}
  \,,
\end{equation}
we can identify~$\tube$ with the Riemannian manifold
$(\tube_0,G)$, where $G\equiv(G_{ij})$ is the metric tensor
induced by the immersion~$\tubemap$,
\ie\/ $G_{ij}:=\tubemap_{,i}\cdot\tubemap_{,j}$.
(Here and in the sequel,
the dot denotes the scalar product in~$\Real^d$
and the comma with an index~$i$ means the partial derivative
w.r.t.~$x_i$, $x\equiv(s,u)\in\Omega_0$.)
Using~(\ref{Frenet}), we find
\begin{equation}\label{metric}
  G =
  \begin{pmatrix}
    h_1 & h_2 & h_3 & \ldots & h_{d-1} & h_d\\
    h_2 & 1 & 0 & \ldots & 0 & 0\\
    h_3 & 0 & 1 & & & 0 \\
    \vdots & & & \ddots & & \vdots \\
    h_{d-1} & & & & 1 & 0 \\
    h_d & 0 & 0 & \ldots & 0 & 1
  \end{pmatrix},
  \quad
  \begin{aligned}
    h_1 & := h^2 + h_\mu h_\mu \,, \\
    h(s,u) & := 1-\kappa_1(s) \, u_2 \,, \\
    h_\mu(s,u) & := -\curvature_{\mu\nu}(s) \, u_\nu \,.
  \end{aligned}
\end{equation}
Furthermore, $|G|:=\det  G=h^2$ which defines through
$d\vol:=h(s,u)\,ds\,du$ the volume element of~$\tube$;
here and in the sequel $du=du_2 \dots du_d$ denotes
the $(d-1)$-dimensional Lebesgue measure in~$\cross$. 

It can be checked by induction that the inverse $G^{-1}\equiv(G^{ij})$
of the metric tensor~(\ref{metric}) satisfies
\begin{equation}\label{inverse.metric}
  G^{-1} = \frac{1}{h^2}
  \begin{pmatrix}
    1 & -h_2 & -h_3 & -h_4 & \ldots & -h_d \\
    -h_2 & h^2+h_2^2 & h_2 h_3 & h_2 h_4 & \ldots & h_2 h_d \\
    -h_3 & h_3 h_2 & h^2+h_3^2 & h_3 h_4 & \ldots & h_3 h_d \\
    \vdots & & & \ddots & \\
    -h_{d-1} & h_{d-1} h_2 & \ldots & & h^2+h_{d-1}^2 & h_{d-1} h_d \\
    -h_d & h_d h_2 & \ldots & & h_d h_{d-1} & h^2+h_d^2
  \end{pmatrix} .
\end{equation}
\begin{Remark}[Low-dimensional examples]\label{Rem.Examples}
When $d=2$, the cross-section~$\omega$ is an interval,  
the curve~$\curve$ has only one curvature~$\kappa:=\kappa_1$
and~$G$ is diagonal with
$$
  h(s,u) = 1-\kappa(s)\,u .
$$
When $d=3$, one finds
$$
  G(\cdot,u) =
  \begin{pmatrix}
    \left(1-\kappa\,u_2\right)^2 + \tau^2\,|u|^2 & -\tau\,u_3 & \tau\,u_2 \\
    -\tau\,u_3 & 1 & 0 \\
    \tau\,u_2  & 0 & 1
  \end{pmatrix} ,
$$
where $\kappa:=\kappa_1$ and $\tau:=\kappa_2$ denote the curvature
and torsion of~$\curve$, respectively.
\end{Remark}
\begin{Remark}[On the assumption~$(\ref{Ass.Basic})$]\label{Rem.basic}
Let $|u|:=\sqrt{u_\mu u_\mu}$ and define
\begin{equation*}
  a:=\sup_{u\in\cross} |u| .
\end{equation*}
By virtue of the inverse function theorem,
$\tubemapalt$ is a local $\Smooth^1$-diffeo\-morphism
provided~$h$ does not vanish on~$\tube_0$.
It becomes a global diffeomorphism if it
is required to be injective in addition.
Hence, (\ref{Ass.Basic})~holds true provided
\begin{Assumption}{\HBasic} 
\begin{itemize}
\item[\emph{(i)}]
$\kappa_1\in\sinf(I)$ and
$a\,\|\kappa_1\|_\infty < 1$\,, 
\item[\emph{(ii)}]
$\tube$ does not overlap itself\,, 
\end{itemize}
\end{Assumption}
which we shall assume henceforth.
Let us point out two facts.
First, if~$\overline{\curve(I)}$ were a compact embedded curve, 
then the condition~(ii) could always be achieved for~$a$ sufficiently small.
Second, we do not need to assume the condition~(ii)
if we consider $(\tube_0,G)$ as an abstract Riemannian manifold
where only the curve~$\curve$ is embedded in~$\Real^d$.
\end{Remark}

For further purposes, we introduce
$$
  \cross^*:=\{u\in\Real^{d-1}|\,(-u_2,u_3,\dots,u_d)\in\cross\}
  \,,
$$
\ie~the mirror image of~$\cross$
w.r.t.~the hyperplane $\{u\in\Real^{d-1}|\,u_2=0\}$.
 
\subsection{The Laplacian}
Introducing the unitary transformation
$
  \Psi \mapsto \Psi\circ\tubemapalt,
$
we may identify the Hilbert space $\sii(\tube)$
with $\Hilbert:=\sii(\tube_0,d\vol)$ and
the Laplacian~(\ref{Laplacian}) with
the self-adjoint operator~$H$ associated with the quadratic
form~$Q$ on~$\Hilbert$ defined by
\begin{align}\label{form}
  Q[\Psi] &:=
  \int_{\tube_0} \overline{\Psi_{,i}(s,u)}\,G^{ij}(s,u)\,\Psi_{,j}(s,u)
  \ h(s,u)\,ds\,du \,,
  \\
  \Psi \in \Dom Q &:=
  \left\{\Psi\in\Sobi(\tube_0,d\vol) | \
  \Psi(s,u)=0
  \quad \textrm{for a.e.} \ (s,u) \in I \times \partial \cross
  \right\} \,.
  \nonumber
\end{align}
Here $\Psi(x)$ for $x\in\partial\tube_0$
means the corresponding trace of the function~$\Psi$ on the boundary.
 
We have
\begin{equation*} 
  H 
  =-|G|^{-\demi}\partial_i |G|^\demi G^{ij} \partial_j
  \,,
\end{equation*}
which is a general expression for the Laplace-Beltrami operator
in a manifold equipped with a metric~$G$.
However, we stress that the equality must be understood
in the form sense if~$\kappa_i$ are not differentiable
(which is the case we are particularly concerned to deal with in this paper).

\section{An intermediate lower bound}\label{Sec.Inter}
In this section, we derive an intemediate lower bound 
to the spectral threshold of~$-\Delta$
which is crucial for the proof of Theorem~\ref{Thm.bound}. 

It is worth to notice that one has the decomposition
\begin{equation}\label{decomposition}
  G^{-1} = \diag(h^{-2},1,\dots,1) + h^{-2} \, \mathcal{T}
  \,,
\end{equation}
where the matrix~$\mathcal{T}$ depends 
on the higher curvatures~$\kappa_\mu$, but not on~$\kappa_1$,
in such a way that $\mathcal{T}=0$ if $\kappa_\mu=0$.
Hence, if the reference curve~$\curve$ is planar (\ie~$\kappa_\mu=0$)
then the norm of a covector $\xi\in T_{(s,u)}^*\tube_0$ 
w.r.t.\ the metric~$G$ is clearly estimated from below by 
the norm of its projection to $T_u^*\cross$
w.r.t.\ the Euclidean norm, \ie\/
$
  \xi_i G^{ij} \xi_j \geq \xi_\mu \xi_\mu 
$.
An important observation 
is that this property is not influenced 
by the presence of higher curvatures: 
\begin{Lemma}\label{Lemma.positive}
One has
$$
  G^{-1} \geq
  \diag(0,1,\dots,1) 
$$
in the matrix-inequality sense.
\end{Lemma}
\begin{proof}
In view of~(\ref{inverse.metric}) and~(\ref{decomposition}),
one has 
$
  G^{-1} - \diag(0,1,\dots,1) 
  = h^{-2} A
$
where
$
  A := \diag(1,0,\dots,0)+\mathcal{T}
$
is positive definite since
\begin{equation*}
  \xi_i A_{ij} \xi_j
  \equiv \xi_1^2 - 2 \, \xi_1 h_\mu \xi_\mu 
  +  (h_\mu \xi_\mu)^2 
  = \left(-\xi_1+h_\mu\xi_\mu\right)^2 \geq 0
\end{equation*}
for any $\xi\in\Real^d$.
\end{proof}

Lemma~\ref{Lemma.positive} has the following crucial corollary.
\begin{Theorem}\label{Thm.bound.begin}
Suppose the assumptions $\HFrenet$ and $\HBasic$ are satisfied.
Then
\begin{equation*} 
  \inf\sigma(-\Delta) \geq
  \inf_{s \in I} 
  \lambda_0\big(\kappa_1(s)\big)
  \,,
\end{equation*}
where
\begin{equation}\label{threshold}
  \lambda_0(\kappa) :=
  \inf_{\psi\in\sobi(\cross)}
  \frac{\int_{\cross}
  \overline{\psi_{,\mu}(u)} \, \psi_{,\mu}(u) \,
  (1-\kappa\,u_2)\,du}
  {\int_{\cross} |\psi(u)|^2 \, (1-\kappa\,u_2)\,du} \,.
\end{equation}
\end{Theorem}
\begin{proof} 
The definition of the form~(\ref{form}),
Lemma~\ref{Lemma.positive} and~(\ref{threshold}) yield
\begin{align*}
  Q[\Psi]
&\geq \int_I ds \int_{\cross} du \
  \overline{\Psi_{,\mu}(s,u)} \, \Psi_{,\mu}(s,u) \,
  \left(1-\kappa_1(s)\,u_2\right) \\
&\geq \int_I ds \ \lambda_0\!\left(\kappa_1(s)\right )
  \int_{\cross} du \ |\Psi(s,u)|^2 \, \left(1-\kappa_1(s)\,u_2\right) \\
&\geq
  \inf_{s \in I} \lambda_0\big(\kappa_1(s)\big) 
  \int_I ds \int_{\cross} du \ |\Psi(s,u)|^2 \, \left(1-\kappa_1(s)\,u_2\right) \\
&\equiv
  \inf_{s \in I} \lambda_0\big(\kappa_1(s)\big) \, \|\Psi\|_\Hilbert^2
\end{align*}
for any $\Psi\in\Dom Q$.
\end{proof}
%
  
\section{Toroidal segments}\label{Sec.Torus}
In this section, we give a geometrical meaning
to the quantity~(\ref{threshold}) and examine its properties,
which then yield Theorem~\ref{Thm.bound}
as a consequence of Theorem~\ref{Thm.bound.begin}.
In particular, the monotonicity properties of 
Proposition~\ref{Prop.Torus.monotonicity} below 
establish the bound~(\ref{bound}) of Theorem~\ref{Thm.bound},
while the uniform lower bound follows from Proposition~\ref{Prop.FK} below. 

Consider now the situation when $I$~is bounded,
$\kappa:=\kappa_1$ is constant
and all $\kappa_\mu = 0$, \ie~$\curve$ is either a circular segment
of length~$|I|$ and radius~$1/|\kappa|$ if~$\kappa\not=0$
or a straight line of length~$|I|$ if~$\kappa=0$.
The assumption~$\HBasic$ holds true provided  
\begin{equation}\label{Ass.torus}
  a \, |\kappa| < 1
  \qquad\textrm{and}\qquad
  |\kappa| \leq 2\pi/|I|
  \,.
\end{equation}
If $\kappa=\pm 2\pi/|I|$, then~$\curve$ is a circle with one point
removed and~$\tube$ is a torus of cross-section~$\omega$ about it
(more precisely, depending on the sign of~$\kappa$,
$\Omega$ can be identified either with
$
  (\Circle\times\cross)\setminus(\{0\}\times\cross)
$
or
$
  (\Circle\times\cross^*)\setminus(\{0\}\times\cross^*)
$,
where~$\Circle$ stands for the one-dimensional sphere
of radius~$1/|\kappa|$).
 
Let~$H^\kappa$ denote the operator associated with~(\ref{form})
in this constant case.
The spectrum of~$H^\kappa$ consists of discrete eigenvalues
which we denote by
$$
  \lambda_0(\kappa,|I|) < \lambda_1(\kappa,|I|) \leq
  \dots \leq \lambda_n(\kappa,|I|) \leq \dots 
  \,,
$$
where the first one is positive.
Since $\curvature_{\mu\nu}=0$ and~$\kappa_1$ is constant, 
the metric~(\ref{metric}) is diagonal and independend
of the ``angular'' variable~$s$.
Consequently, the coefficients of~$H^\kappa$ do not depend on~$s$ either
and the Laplacian can be decomposed
w.r.t.\ the angular momentum subspaces represented by 
the eigenfunctions of~$-\Delta_N^I$,
\ie~the Neumann Laplacian on~$\sii(I)$.
\begin{Lemma}\label{Lemma.decomposition}
Let~$\phi_n$, $n\in\Nat$, denote the normalised eigenfunction corresponding 
to the $(n+1)^\mathrm{th}$ eigenvalue $E_n := (\pi/|I|)^2 n^2$
of~$-\Delta_N^I$.
Then $H^\kappa$ is unitarily equivalent to the direct sum
$
  \bigoplus_{n\in\Nat} H_n^\kappa \,,
$
where each $H_n^\kappa$ acts on
$
  \{\phi_n\} \otimes \sii\big(\cross,(1-\kappa\,u_2)\,du\big)
$
and it is defined in the form sense by
$$
  H_n^\kappa 
  := \frac{E_n}{(1-\kappa\,u_2)^2}
  - \frac{1}{1-\kappa\,u_2} \partial_\mu (1-\kappa\,u_2) \partial_\mu
  \,, \qquad
  \Dom (H_n^\kappa)^\demi 
  := \{\phi_n\} \otimes \sobi(\cross) .
$$
Furthermore, each~$H_n^\kappa$ is unitarily equivalent
to the operator~$\hat{H}_n^\kappa$ on
$
  \{\phi_n\} \otimes \sii(\cross)
$
defined in the form sense by
$$
  \hat{H}_n^\kappa :=
  \Id\otimes(-\Delta_D^{\cross}) + V_n^\kappa
  \,,\qquad
  \Dom (\hat{H}_n^\kappa)^\demi := \{\phi_n\} \otimes \sobi(\cross) ,
$$
where
\begin{equation}\label{potential}
  V_n^\kappa(u_2)
  := \frac{E_n-\kappa^2/4}{(1-\kappa\,u_2)^2}  
\end{equation}
and $-\Delta_D^\cross$ denotes the Dirichlet Laplacian on~$\sii(\omega)$.
\end{Lemma}
\begin{proof}
Since~$\kappa$ is constant, $h(s,u)$ is independent of~$s$
and we have the following natural isomorphisms
\begin{eqnarray*}
  \Hilbert
  &\simeq& \sii(I) \otimes \sii(\cross,(1-\kappa\,u_2)\,du),
  \\
  \Dom Q 
  &\simeq& \Dom(-\Delta_N^I)^\demi
  \otimes \sobi(\cross,(1-\kappa\,u_2)\,du).
\end{eqnarray*}
Since the family~$\{\phi_n\}_{n\in\Nat}$ 
forms a complete orthonormal basis in~$\sii(I)$,
the Hilbert space~$\Hilbert$ admits a direct sum decomposition
$
  \Hilbert = \bigoplus_{n\in\Nat} \Hilbert_n ,
$
where
$
  \Hilbert_n := \{\phi_n\} \otimes \sii(\cross,(1-\kappa\,u_2)\,du).
$
Noticing that the spaces $\sobi(\cross,(1-\kappa\,u_2)\,du)$
and $\sobi(\cross)$ can be identified as sets,
we arrive at the first claim of the Lemma because 
$
  Q[\psi] = \left(\psi,H_n^\kappa\psi\right)
$
for any $\psi\in\Dom (H_n^\kappa)^\demi$. 
The second claim follows by means of the transformation
$
  \psi \mapsto (1-\kappa\,u_2)^\demi\psi ,
$
which is unitary from $\Hilbert_n$ 
to $\{\phi_n\}\otimes\sii(\cross)$
and leaves invariant $\Dom (H_n^\kappa)^\demi$.
\end{proof}

Let us recall that the spectrum of~$-\Delta_D^\cross$ consists 
of discrete eigenvalues which we denote by
$$
  \mu_0 < \mu_1 \leq \dots \leq \mu_n \leq \dots,
  \qquad n\in\Nat 
  \,,
$$
where the lowest eigenvalue~$\mu_0$ is positive.
  
Lemma~\ref{Lemma.decomposition} is useful in order to investigate
the spectrum of~$H^\kappa$. Here we employ it just to establish 
some properties of the first eigenvalue. 
Since the spectrum of a direct sum of self-adjoint operators
is given by the sum of the individual spectra,
\cf~\cite[Corol. of Thm.~VIII.33]{RS1},
$\lambda_0(\kappa,|I|)$ is just
the first eigenvalue of~$\hat{H}_0^\kappa$ (and~$H_0^\kappa$). 

The first observation is that 
$\lambda_0(\kappa,|I|)$ \emph{does not depend on}~$|I|$
because \mbox{$E_0=0$}. 
This fact is easy to understand because $\lambda_0(\kappa,|I|)$,
with~$\kappa\not=0$, is nothing else than the first eigenvalue 
of the Dirichlet Laplacian in a torus of cross section~$\cross$
and it is known that the corresponding eigenfunction
is invariant w.r.t.\ the rotations around the point of symmetry
($\lambda_0(0,|I|)$ is the spectral threshold
of an infinite straight tube of cross-section~$\cross$
which is equal to~$\mu_0$).
In fact, as a direct consequence of a variational formula 
for the lowest eigenvalue of~$H_0^\kappa$, 
we get the identity 
\begin{equation}
  \lambda_0(\kappa,|I|) = \lambda_0(\kappa)  
  \,,
\end{equation}
where the latter is given by~(\ref{threshold}).

Henceforth, we consider $\kappa\mapsto\lambda_0(\kappa)$
as a function on $(-1/a,1/a)$ and examine its properties
by means of the second part of Lemma~\ref{Lemma.decomposition}
(an alternative, equivalent, approach is to make the change 
of trial function $\psi\mapsto(1-\kappa\,u_2)^{-\demi}\psi$
directly in~(\ref{threshold}), which makes the denominator 
of the Rayleigh quotient independent of~$\kappa$,
while the potential~$V_0^\kappa$ appears in the numerator). 

The following result together with Theorem~\ref{Thm.bound.begin}
establishes the lower bound~(\ref{bound}) of Theorem~\ref{Thm.bound}.
\begin{Proposition}[Monotonicity]\label{Prop.Torus.monotonicity}
The function $\kappa \mapsto \lambda_0(\kappa)$ is
\begin{itemize}
\item[\emph{(i)}]
continuous on $ (-1/a,1/a) ;$
\item[\emph{(ii)}]
increasing on $\left(-1/a,0\right] ;$
\item[\emph{(iii)}]
decreasing on $\left[0,1/a\right)$.
\end{itemize}
\end{Proposition}
\begin{proof}
\ad~(i).
This is immediate from the minimax principle
applied to~$\hat{H}_0^\both$.
\\
\ad~(ii) and~(iii).
Calculating
$$
  \frac{\partial V_0^\kappa}{\partial\kappa} (u_2)
  = -\frac{\kappa}{2(1-\kappa\,u_2)^3},
$$
we see that the potential~(\ref{potential}) as a function of~$\kappa$ is
increasing for $\kappa \leq 0$
and decreasing for $\kappa \geq 0$. 
The claim then follows easily by the minimax principle.
\end{proof}

The following result follows from the fact that 
the operator~$\hat{H}_0^\kappa$ is invariant
w.r.t.\ the simultaneous change
$\kappa \mapsto -\kappa$ and~$u_2 \mapsto -u_2$.
\begin{Proposition}[Symmetry]\label{Prop.symmetry} 
If $\cross=\cross^*$,  
then the function $\kappa \mapsto \lambda_0(\kappa)$
is even on $(-1/a,1/a)$.
\end{Proposition}

We note that~$\mu_0$, as an eigenvalue of the Dirichlet Laplacian, 
has the asymptotics~$\mu_0=\mathcal{O}(a^{-2})$ as~$a \to 0$.
Since one is dealing with Dirichlet boundary conditions 
on $I\times\partial\omega$, one expects  
the same behaviour from~$\lambda_0(\kappa)$.
We derive the following asymptotics.  
\begin{Proposition}[Thin-width asymptotics]\label{Prop.thin}
One has
$$
  \lambda_0(\kappa)
  = \mu_0 -\mbox{$\frac{1}{4}$} \kappa^2 + \mathcal{O}(a)
  \qquad\textrm{as}\quad
  a \to 0.
$$
\end{Proposition}
\begin{proof}
Since
$
  V_0^\kappa(u_2)
  = - \mbox{$\frac{1}{4}$}\kappa^2 + \mathcal{O}(u_2),
$
the result immediately follows by the minimax principle.
\end{proof}

Finally, applying the Faber-Krahn inequality to~$\lambda_0(\kappa)$
with help of Proposition~\ref{Prop.Torus.monotonicity},
one obtains the uniform lower bound of Theorem~\ref{Thm.bound}.
\begin{Proposition}[Uniform bound]\label{Prop.FK}
One has
\begin{equation*} 
  \forall\kappa\in(-1/a,1/a),\qquad
  \lambda_0(\kappa) \geq
  c :=
  \left(
  \frac{|\Sphere^{d-1}|}{d\,|\Sphere^1|\,a\,|\cross|}
  \right)^\frac{2}{d}
  j_{(d-2)/2,1}^2 
  \,,
\end{equation*}
where~$j_{(d-2)/2,1}$
denotes the first zero of the Bessel function~$J_{(d-2)/2}$.
\end{Proposition}
%
 
\section{Conclusions}\label{Sec.Disc}
%
The main goal of this paper was to derive a lower bound 
to the spectral threshold of the Laplacian~(\ref{Laplacian})
in curved tubes~(\ref{tube}).  
Our Theorem~\ref{Thm.bound} states that this bound 
is given by~$\lambda_0(\kappa)$, 
\ie~the lowest eigenvalue of the Dirichlet Laplacian
in a torus of curvature~$\kappa$, 
with~$\kappa$ being determined uniquely by the first curvature 
of the reference curve and the tube cross-section.  
It follows from Section~\ref{Sec.Torus} that 
$\kappa\mapsto\lambda_0(\pm\kappa)$ is a decreasing function  
(\cf~Proposition~\ref{Prop.Torus.monotonicity}),
\ie\/ \emph{bending diminishes the lower bound}
(see also Proposition~\ref{Prop.thin}). 
Another interesting observation is that the lower bound 
does not depend on higher curvatures of the reference curve
(technically, this is a consequence of Lemma~\ref{Lemma.positive}),
\ie\/ \emph{twisting does not diminish the lower bound}. 

We note that Proposition~\ref{Prop.symmetry} yields
$
  \inf\sigma(-\Delta) \geq
  \lambda_0(\|\kappa_1\|_\infty)
$
provided $\cross=\cross^*$,
and Proposition~\ref{Prop.thin} implies asymptotics
of the lower bound for thin tubes.
 
It follows immediately from the minimax principle that 
the lower bound of Theorem~\ref{Thm.bound} also applies 
to other boundary conditions 
imposed on~$\tubemap((\partial I)\times\cross)$,
\eg, Dirichlet, Robin, periodic, \etc.

Adapting the approach of Section~\ref{Sec.Torus}
to the case of Dirichlet boundary conditions 
imposed everywhere on~$\partial\Omega$,
one reveals interesting isoperimetric inequalities
for the first eigenvalue, denoted here by~$\lambda_0^D(\kappa,|I|)$, 
of the Laplacian in a toroidal segment~$\Omega$ 
of curvature~$\kappa$, length~$|I|$ and cross-section~$\cross$.
In particular, $\kappa\mapsto\lambda_0^D(\kappa,|I|)$
attains its minimum for $\kappa=\pm 2\pi/|I|$,
\ie\/ when~$\Omega$ is the whole torus
with a supplementary Dirichlet condition imposed
on a transverse cross-section~$\cross$
(\cf~the beginning of Section~\ref{Sec.Torus}).
This minimum is equal to the first eigenvalue~$\mu_0$ 
of the Dirichlet Laplacian in~$\cross$
and therefore it depends neither on~$|I|$,
nor on the rotations of~$\cross$.
At the same time, it can be shown that 
$\kappa\mapsto\lambda_0^D(\pm\kappa,|I|)$
is decreasing on the interval 
$[4a\pi^2/|I|^2,2\pi/|I|]$.  
Furthermore, if~$d=2$, one can modify
the proof of Theorem~2 in~\cite{Laugesen1} 
and show that the maximum is attained for~$\kappa=0$,
\ie\/ when~$\Omega$ is a rectangle. 
An open problem is to prove (or disprove)
the monotonicity on $[0,4a\pi^2/|I|^2]$.

Let us also mention that the lower bound
of Theorem~\ref{Thm.bound} is optimal in the sense
that the equality is achieved for a tube geometry (a torus or a straight tube).
However, the question about an optimal lower bound
in an unbounded curved tube is more difficult and remains open.

The hypothesis~$\HBasic$ was discussed in Remark~\ref{Rem.basic}. 
As mentioned in Remark~\ref{Rem.Frenet}, 
our hypothesis~$\HFrenet$ allows us to consider 
some curves which do not possess a distinguished Frenet frame.
However, there still exist curves for which the hypothesis~$\HFrenet$ fails;
see~\cite[Chap.~1, p.~34]{Spivak2} for an example 
of such a ($\Smooth^\infty$-smooth but not analytic) curve in~$\Real^3$.
Without going into details, 
let us only mention that the hypothesis~$\HFrenet$ 
is not necessary for the lower bound~(\ref{bound}) to hold. 
For instance, using a Neumann bracketing argument,
it suffices to assume that the hypothesis~$\HFrenet$ is satisfied ``piece-wise";
this may happen if there are isolated points when some of the curvatures vanish.

Let us conclude this paper by comparing 
the result of Theorem~\ref{Thm.bound}
with the lower bound established in~\cite{AE}
in the situation when~$I=\Real$, $d=2,3$,
the cross-section was circular
and the discrete spectrum of~$-\Delta$ was not empty but finite.
The results of~\cite{AE} read as
\begin{equation*}
  \inf\sigma(-\Delta)
  \geq
  \begin{cases}
    3^{1-N}
    \left(j_{0,1}/j_{1,1}\right)^2 \mu_0
    \ \approx\
    3^{1-N} \, 0.3939 \, \mu_0
    & \textrm{if}\quad d=2,
    \\
    \left(\pi/j_{3/2,1}\right)^2 \mu_0
    \ \approx\
    0.4888 \, \mu_0
    & \textrm{if}\quad d=3, \ N=1.
  \end{cases}
\end{equation*}
where~$N$ is the number of discrete eigenvalues (counting multiplicity).
Our uniform lower bound given by Proposition~\ref{Prop.FK} 
can be written as
\begin{equation*}
  \inf\sigma(-\Delta)
  \geq
  \begin{cases}
    \left(j_{0,1}/\pi\right)^2 \mu_0
    \ \approx\
    0.5860 \, \mu_0
    & \textrm{if}\quad d=2,
    \\
    (2/(3\pi))^{2/3}
    \left(j_{1/2,1}/j_{0,1}\right)^2 \mu_0
    \ \approx\
    0.6072 \, \mu_0
    & \textrm{if}\quad d=3,
  \end{cases}
\end{equation*}
which is evidently better and applies to tubes with an infinite number,
or without any, discrete eigenvalues, too;
we also emphasise that we have compared the results of~\cite{AE}
with a crude bound of Proposition~\ref{Prop.FK},
a better bound to $\inf\sigma(-\Delta)$
is contained in~(\ref{bound}) of our Theorem~\ref{Thm.bound}.

\section*{Acknowledgements}
\addcontentsline{toc}{section}{Acknowledgements}
The authors thank the referee 
for helpful suggestions which have improved the presentation. 
This work was partially supported by FCT/POCTI/FEDER, Portugal,
and GA AS CR grant IAA 1048101.

%
%
\addcontentsline{toc}{section}{References}
\bibliography{bib}
\bibliographystyle{amsalpha}
\end{document}